\documentclass{shnature}

\usepackage{color}
\usepackage{graphics} 
\usepackage{graphicx} 
\usepackage{subfigure}

\newcounter{eqcounter}
\newenvironment{smequation}{%
\addtocounter{equation}{-1}
\refstepcounter{eqcounter}

\begin{equation}}
{\end{equation}}

\newcounter{figcounter}
\newenvironment{smfigure}{%
\addtocounter{figure}{-1}
\refstepcounter{figcounter}

\begin{figure}}
{\end{figure}}

\title{Implementation of a canonical phase measurement with quantum feedback}

\author{Leigh S. Martin,$^{1,2\ast}$ William P. Livingston,$^{1,2}$ Shay Hacohen-Gourgy,$^{1,2,3}$\\ Howard M. Wiseman,$^{4}$ Irfan Siddiqi$^{1,2}$}

\begin{document}

\maketitle

\begin{affiliations}
 \item Center for Quantum Coherent Science, Berkeley, CA 94720 USA
 \item Department of Physics, University of California, Berkeley, CA 94720 USA
 \item Department of Physics, Technion - Israel Institute of Technology, Haifa 32000 Israel
 \item Centre for Quantum Computation and Communication Technology (Australian Research Council), Centre for Quantum Dynamics, Griffith University, Nathan, QLD 4111 Australia
\end{affiliations}

\begin{abstract}
Much of modern metrology and communication technology encodes information in electromagnetic waves, typically as an amplitude or phase. While current hardware can perform near-ideal measurements of photon number or field amplitude, to date no device exists that can even in principle perform an ideal phase measurement. In this work, we implement a single-shot canonical phase measurement on a one-photon wave packet, which surpasses the current standard of heterodyne detection and is optimal for single-shot phase estimation. By applying quantum feedback to a Josephson parametric amplifier, our system adaptively changes its measurement basis during photon arrival and allows us to validate the detector's performance by tracking the quantum state of the photon source. These results provide an important capability for optical quantum computing, and demonstrate that quantum feedback can both enhance the precision of a detector and enable it to measure new classes of physical observables.
\end{abstract}

From FM radio communication to gravitational wave detection, low-noise measurement of phase is integral to much of modern communication and sensing technology. To understand the limits of such applications, one must look at their quantum mechanical descriptions, which sets an ultimate bound. Although no quantum mechanical operator corresponds to phase\cite{busch2001number}, one can nevertheless define an ideal measurement basis $|\theta\rangle = \sum_{n=0}^\infty e^{i n \theta}|n\rangle$ that yields a canonical phase measurement\cite{London1926,susskind1964phase}, where $|n\rangle$ is a quantum state of $n$ photons, and $\theta$ parameterizes the measurement outcome. As $|\theta\rangle$ contains a uniform superposition of all photon-number states, a measurement outcome in this basis yields no information about photon number. In this sense, phase and photon number are complementary variables\cite{busch2001number}; there is a direct trade-off between measuring one and measuring the other, much like the well-known Heisenberg incompatibility between position and momentum. This trade-off becomes particularly important at low photon numbers, where intrinsic quantum mechanical uncertainty of the transmitted state becomes significant.

In the absence of an instrument capable of implementing a canonical phase measurement, heterodyne detection, in which one measures a rapidly varying quadrature of the input, serves as the standard technique for estimating the phase of an unknown signal. Several schemes have surpassed heterodyne detection 
\cite{yonezawa2012phasetracking,wheatley2010adaptive,armen2002adaptive,iwasawa2013mirrormotion}, 
 however these protocols also acquire undesired photon number information, and thus cannot reach the quantum limit or implement a canonical phase measurement.
In this work, we implement a feedback-controlled quantum-limited amplifier which dynamically updates its amplitude measurement in response to the incident field. When the system continuously optimizes this measurement basis for phase sensitivity, it implements a canonical phase measurement on an incoming single-microwave-photon state\cite{wiseman1995adaptive}. We verify implementation of a canonical phase measurement using the entanglement between the emitted photon and its source, which allows us to confirm that acquisition of photon number information is suppressed. The system surpasses heterodyne detection by $15\pm 2$\%. 



\begin{figure}
{\includegraphics[width = 100mm]{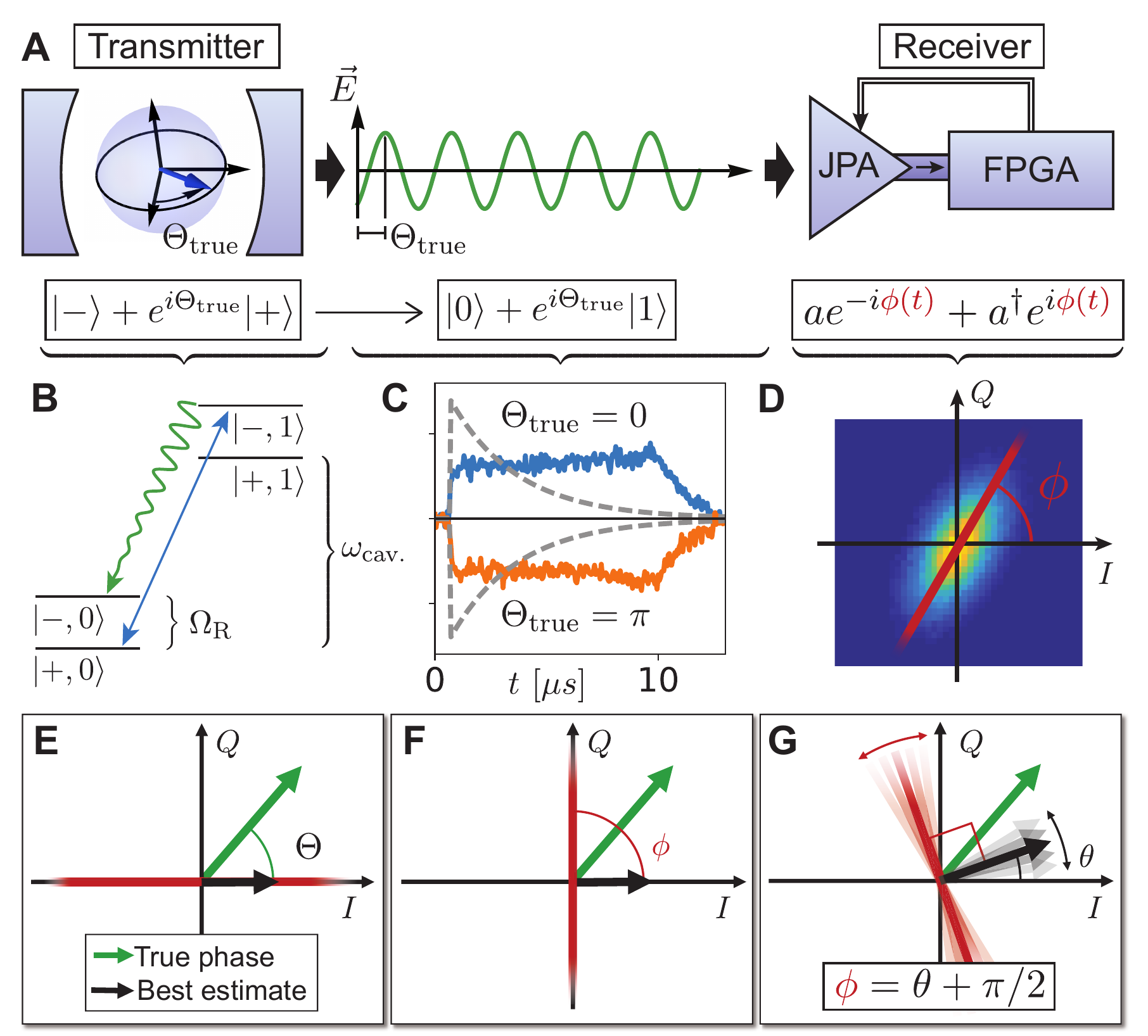}} 
\caption{
Experimental implementation.
(\textbf{A}) Atom in a cavity, with phase $\Theta_\mathrm{true}$ encoded into its dipole moment. The atom decays and emits a photon into a 1D waveguide with phase encoded into the electric field as shown. The JPA receives the photon and measures an amplitude quadrature selected by the FGPA.
(\textbf{B}) Sideband cooling scheme to emit photon. Sideband converts a qubit excitation to a cavity excitation, which is then emitted as a single photon at the cavity frequency.
(\textbf{C}) Measured mode shape (E-field envelope) of emitted photon. Dashed line shows mode shape if constant cooling rate were used instead.
(\textbf{D}) Output of JPA. Signal is amplified along measurement axis $\phi$ and squeezed along the other. 
(\textbf{E-G}) Estimating and tracking state by changing measurement basis. Receiver attempts to maintain the phase measurement condition $\phi=\theta+\pi/2$. See text for details.
}
\label{fig:ExperimentalSetup}
\end{figure}

%
%
As shown in Fig. \ref{fig:ExperimentalSetup}A, our system consists of a transmitter, which encodes a variable $\Theta_\mathrm{true}$ into the phase of a single-photon electromagnetic signal, and a receiver, which uses a continuous feedback protocol to guess this phase in a single shot using an adaptive feedback protocol. A superconducting transmon qubit\cite{koch2007transmon} embedded in a 3D aluminum cavity acts as the transmitter. We use coherent bath engineering \cite{murch2012cavity} of this artificial atom to generate our photonic state, which yields more process control than direct spontaneous decay. To implement this scheme, we Rabi drive our qubit at $\Omega_\mathrm{R}/2\pi = 20$ MHz, which creates an effective low-frequency qubit. Simultaneously, we apply a cavity sideband at $\omega_\mathrm{cav.}+\Omega_\mathrm{R}$, where $\omega_\mathrm{cav.}$ is the cavity resonance frequency. As shown in Fig. \ref{fig:ExperimentalSetup}B, the sideband drives a transition from the $|+,0\rangle$ state to $|-,1\rangle$ state, where $|\pm\rangle \equiv (|e\rangle \pm i|g\rangle)/\sqrt{2}$ are the dressed states of qubit under driving and $0,1$ count the number of photons in the cavity. The cavity then decays, emitting a photon and leaving the system in the $|-,0\rangle$ state, which is not affected by the sideband. We ensure that the cavity decay rate is fast compared to the sideband-induced coupling, so that the qubit's effective decay rate from $|+\rangle$ to $|-\rangle$ is limited by the sideband amplitude. By modulating the sideband amplitude during photon emission\cite{SupplementaryMaterials}, we generate a photon with a flat modeshape (Fig. \ref{fig:ExperimentalSetup}C), which greatly ameliorates the detrimental effects of feedback delay at the receiver\cite{pozza2015deterministic}. To encode the phase $\Theta_\mathrm{true}$, we prepare the qubit in a superposition state of the form $(|-\rangle+e^{i\Theta_\mathrm{true}}|+\rangle)/\sqrt{2}$, which decays by emitting the photonic state $(|0\rangle + e^{i\Theta_\mathrm{true}}|1\rangle)/\sqrt{2}$.

Our receiver consists of a Josephson parametric amplifier (JPA)\cite{castellanos2008amplification} pumped at twice its resonance frequency by a field-programmable gate array (FPGA), which serves as a classical feedback controller (Fig \ref{fig:ExperimentalSetup}A). To maintain high measurement bandwidth for quantum feedback, we operate the JPA at a relatively low gain of $6$ dB, which yields a gain bandwidth of $45$ MHz, and follow it with a traveling wave parametric amplifier\cite{macklin2015twpa} (not shown) to boost the signal strength and maintain a quantum efficiency of $\eta = 0.4$. The JPA measures field amplitude via the quantum mechanical quadrature operator $a e^{-i\phi(t)} + a^\dagger e^{i\phi(t)}$, where $a$ is the quantum mechanical annihilation operator of the incident field and $\phi(t)$ is the instantaneous phase of the parametric pump. 

To perform a canonical phase measurement on the incident field, the feedback controller continuously adapts the measurement axis $\phi(t)$ as the photon arrives at the receiver\cite{wiseman1995adaptive}. The measurement axis is chosen to maximize the acquisition of phase information as follows. Before the photon reaches the JPA, the receiver has no information and therefore chooses $\phi$ arbitrarily. Upon arrival of a portion of the photon, the JPA detects a small positive (or negative) fluctuation, which then informs the system that the true phase is likely oriented along (or opposite) the measurement axis (Fig. \ref{fig:ExperimentalSetup}E). At this point, any further measurement in this basis interrogates the amplitude of the incident field and thus yields undesired photon number information. Ideally, the system would then rotate the measurement axis by 90 degrees (Fig. \ref{fig:ExperimentalSetup}F), so that a small deviation between the current best estimate of the phase $\theta(t)$ and the true phase $\Theta_\mathrm{true}$ would be detectable as a positive or negative fluctuation in the signal. As the photon continues to arrive, the feedback controller gains more information and updates the phase $\phi(t)$ to maximize sensitivity to phase (Fig. \ref{fig:ExperimentalSetup}G). If the phase measurement condition $\phi(t) = \theta(t) + \pi/2$ is maintained at all times, then the system acquires no photon number information and implements a canonical phase measurement.



%
%
To track the best estimate of the phase, the feedback controller must continuously update its best guess of the atom's state based on the measurement signal starting with no prior information \textit{i.e.} it should track the quantum trajectories of the system\cite{Murch:2013ur,campagne2016observing} given an initially maximally mixed state. 
We begin by observing and verifying quantum trajectories for homodyne ($\phi(t)=0$) and heterodyne ($\phi(t) = \omega_\mathrm{het.} t$, $\omega_\mathrm{het.}/2\pi =0.5$ MHz) detection\cite{campagne2016observing}. Example trajectories are plotted in Fig. \ref{fig:BackAction}A,B and tomographically validated in \cite{SupplementaryMaterials}. These data allow us to characterize measurement back-action and check consistency with theory. The stochastic component of the back-action always lies in the plane of the instantaneous measurement basis, as is clear from the homodyne data.


The presence of back-action not only governs how to adapt the measurement axis $\phi(t)$, but also offers a method to independently validate the receiver's implementation of a canonical phase measurement. 
Because an ideal phase measurement acquires maximal phase information and no photon-number information, it maximally disturbs the atomic dipole phase while minimally disturbing the atomic excitation probability. This effect is directly visible in the quantum trajectories, as illustrated conceptually in Fig. \ref{fig:BackAction}C and D. When the measurement axis is aligned with the best estimate of the phase ($\phi = \theta$), the resulting acquisition of amplitude information manifests as a random disturbance of the qubit state along the axis of decay (Fig. \ref{fig:BackAction}C). Conversely, when the phase measurement condition is satisfied ($\phi = \theta+\pi/2$), then only the phase of the qubit state is subject to noise (Fig. \ref{fig:BackAction}D). In this way, we can verify the performance of our receiver by characterizing the dynamics of the transmitter. This capability is uniquely quantum, and arises from entanglement between the atom and its emitted photon.


We show the results of this verification scheme in Fig. \ref{fig:AdaptiveBackaction}. Fig. \ref{fig:AdaptiveBackaction}A shows a single quantum trajectory under adaptivedyne detection, in which $\phi(t)$ is continuously adapted by the feedback controller. Fig. \ref{fig:AdaptiveBackaction}B shows the difference between the ideal quadrature phase and the measured phase, which shows that the feedback controller approximately maintains the phase measurement condition. To interpret the dynamics, we plot the ensemble statistics of the phase back-action as a function of time in Fig. \ref{fig:AdaptiveBackaction}D, with the heterodyne detection case included for comparison. It can be seen that the phase back-action $d\theta$ is significantly larger for adaptivedyne detection. Fig. \ref{fig:AdaptiveBackaction}C shows the ensemble statistics of the state at $t=10 ~\mu$s. As observed in \cite{campagne2016observing}, the quantum trajectories of a decaying atom evolve on a spherical shell that shrinks deterministically to the south pole of the Bloch sphere. Due to the suppression of back-action along the decay axis, adaptivedyne trajectories are further confined, exhibiting something closer to a ring-like structure. This feature presents an unambiguous signal that our system approximately implements a canonical phase measurement. 

A canonical phase measurement should outperform heterodyne detection in estimating the phase $\Theta_\mathrm{true}$. To verify superior performance, we prepare our qubit in one of 8 equally spaced points along the equator of the Bloch sphere. From each shot, the receiver optimally\cite{pozza2015deterministic} estimates the phase of the photon by computing the following quantity
\begin{equation} \label{eq:R}
R = \int_0^T e^{i \phi(t)}\sqrt{u(t)}V(t) dt
\end{equation}
where $u(t)$ is the photon mode shape, and $T$ is the duration of each experimental run and $V(t)$ is the measurement signal read out from the JPA normalized such that its variance is $dt$. The best estimate of the photon's phase in a single shot is given by the complex argument $\theta(T) = \arg(R)$. Fig. \ref{fig:PhaseEstimation}A plots a histograms of this best estimate for adaptivedyne detection, which exhibits the $\cos(\theta-\Theta_\mathrm{true})$ dependence expected theoretically\cite{pozza2015deterministic}.

We compare the performance of adaptivedyne and heterodyne detection by plotting the Holevo variance of each underlying distribution in Fig. \ref{fig:PhaseEstimation}B. We also include data for what we term replay detection, in which $\phi(t)$ from an adaptivedyne shot of the experiment is replayed instead of feeding back based on the current signal. In this way, we can confirm that it is the correlations between $\phi(t)$ and the state that yield enhanced performance, rather than the independent statistics of $\phi(t)$. For additional confirmation, we independently measure the signal-to-noise ratio of our amplifier chain for heterodyne and adaptivedyne detection and verify that it remains the same to well within 1\%\cite{SupplementaryMaterials}. Heterodyne and replay perform equally well, and are both significantly surpassed by a canonical phase measurement implemented via adaptivedyne detection. Adaptivedyne does not reach the quantum limited Holevo variance of 3 due to a combination of loss, qubit decoherence and feedback delay. However from our heterodyne data we infer an adjusted quantum limit given our quantum efficiency and purity of the emitted photon, as well as the hypothetical homodyne limit. The canonical phase measurement comes significantly closer to this adjusted quantum limit than any other scheme, limited almost entirely by feedback delay.

We infer the sensitivity of each scheme to photon-number information from the distributions of $|R|$, which are shown in Fig. \ref{fig:PhaseEstimation}. The distributions for heterodyne and replay are almost identical, while the adaptivedyne histogram is substantially narrower, indicating that the latter is less sensitive to this undesired information\cite{pozza2015deterministic}. 

Several avenues remain for future work. Firstly, we have optimized the system for fair comparison between heterodyne and adaptivedyne detection in order to be sure that the observed improvements arise from feedback alone. A system that is optimized for adaptivedyne could easily yield further improvements. Quantum efficiencies as high as 80\% have been demonstrated in circuit QED\cite{eddins2018transamp}, and similar improvements could be achieved by increasing amplifier gain or adding low-loss or on-chip circulators\cite{kerckhoff2015onchip}. Integrating low-temperature electronics closer to the amplifier could also significantly reduce feedback latency, which would yield immediate gains in the phase estimation efficiency.

Our system has several immediate applications to quantum information and computation. Firstly, the implementation of quantum feedback on a detector is known to allow enhanced readout of superconducting circuits\cite{sarovar2007adaptive}. Furthermore, the ability to perform a canonical phase measurement enables linear-optics preparation of the $|0\rangle+|1\rangle$ photonic state, which is a major experimental challenge of single-rail linear optics quantum computing\cite{ralph2005adaptive}. More broadly, it is known that adaptive measurements are universal\cite{oreshkov2005weak}, meaning that many relatively simple measurement devices augmented with quantum feedback can perform any measurement allowed by quantum mechanics. Thus our extension of a standard amplitude measurement device to an ideal phase measurement represents a more general and exciting direction for future research.

\begin{figure}
{\includegraphics[width = 100mm]{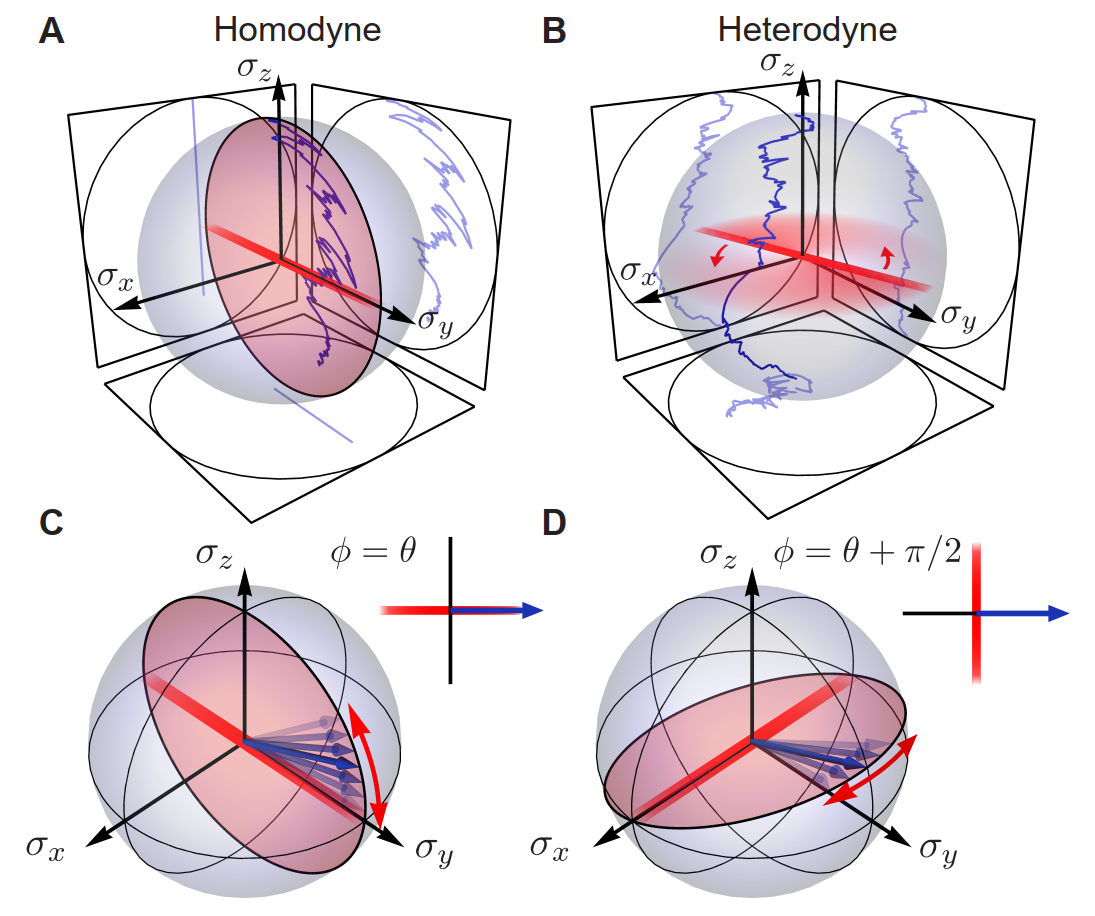}} 
\caption{
Measurement back-action and quantum trajectories. Coordinate axes are chosen so that the atom decays to $\sigma_z=-1$. (\textbf{A}) A single homodyne quantum trajectory ($\phi(t) = 0$). State only propagates in the plane of the measurement axis. (\textbf{B}) A single heterodyne trajectory ($\phi(t) = \omega_\mathrm{het.} t$). The qubit is initialized in $|+\rangle$ for both trajectories. (\textbf{C}) Amplitude back-action, which occurs when the measurement axis (red line) is aligned to the best estimate of the state (blue arrow). (\textbf{D}) Phase back-action, which occurs when the phase measurement condition is satisfied. 
}
\label{fig:BackAction}
\end{figure}

\begin{figure} 
{\includegraphics[width = 100mm]{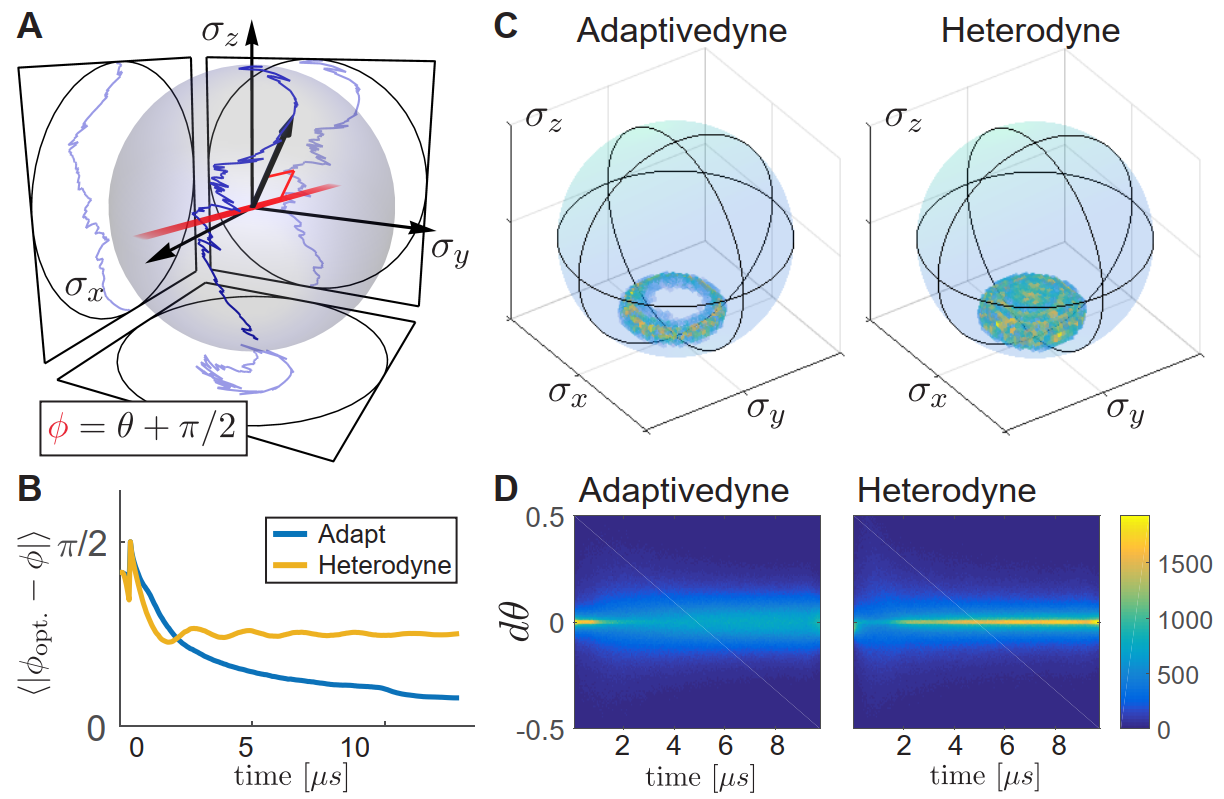}} 
\caption{
Back-action and measurement validation. (\textbf{A}) A single adaptive-dyne quantum trajectory. The red right-angle bracket emphasizes orthogonality between the measurement axis and the state. (\textbf{B}) Quality of tracking for heterodyne and adaptivedyne, where $\phi_\mathrm{opt.} = \theta(t)+\pi/2$. Adaptivedyne significantly outperforms the heterodyne and comes close to the ideal phase by $T=13 \mu s$. The difference $\phi_\mathrm{opt.}-\phi$ is cut to lie on the interval $[-\pi/2,\pi/2]$. (\textbf{C}) Distribution of trajectories at $t=10 \mu s$. Due to suppression of photon-number back-action, adaptivedyne trajectories cluster in a ring at late times.
(\textbf{D}) Statistics of the phase back-action $d\theta$ for adaptivedyne and heterodyne. 
On average, the phase back-action is significantly larger for adaptivedyne. 
}
\label{fig:AdaptiveBackaction}
\end{figure}

\begin{figure} 
{\includegraphics[width = 120mm]{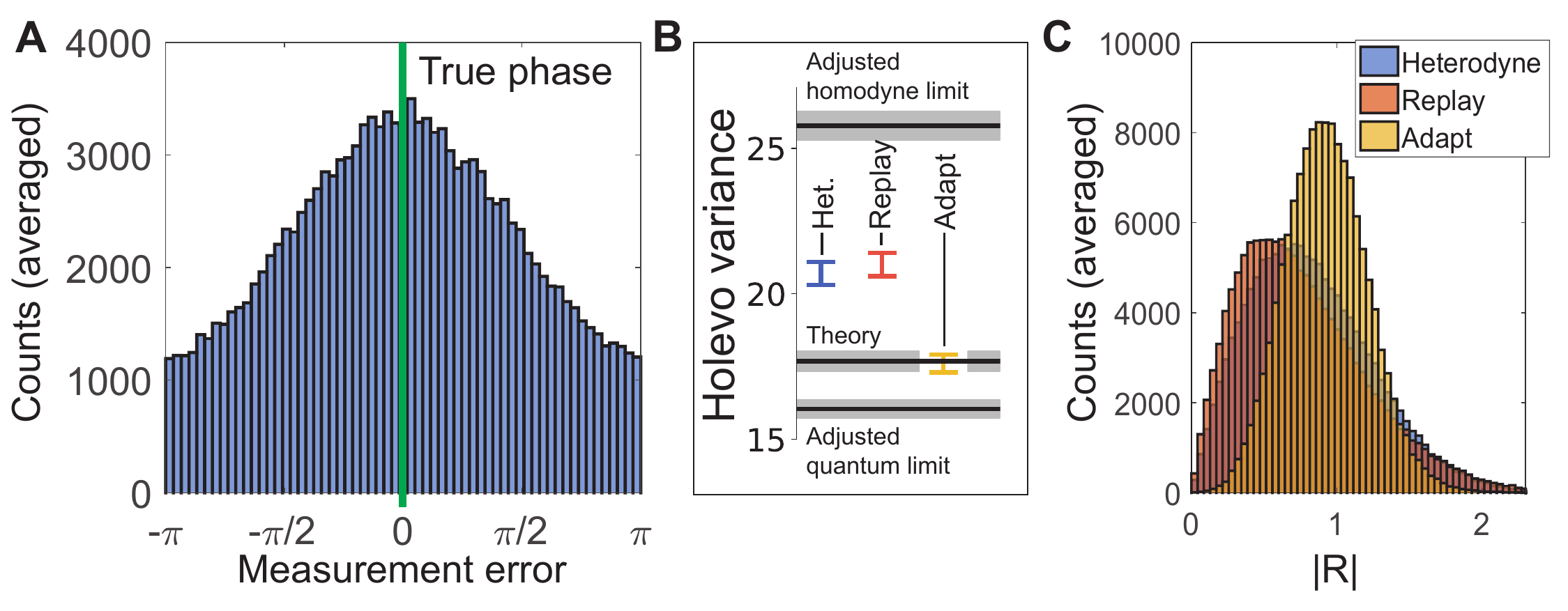}}
\caption{
Phase-estimation performance. (\textbf{A}) Histogram of the difference between the measurement outcome and the true phase \textit{i.e.} $\theta(T)-\Theta_\mathrm{true}$ (\textbf{B}) Performance is evaluated by computing the Holevo variance of this distribution. Quantum limit (bottom black line) homodyne limit (top black line) and absolute theory prediction based on feedback delay are inferred from the performance of heterodyne, with corresponding error bars shown as gray rectangles. (\textbf{C}) Distribution of the amplitude information. The distribution is significantly narrower for adaptivedyne, indicating suppression of this information channel. 
}
\label{fig:PhaseEstimation}
\end{figure}

\begin{methods}
\subsection{Experimental details and devices}

\textbf{Transmon qubit, cavity and measurement chain} A full wiring diagram for our low-temperature and room-temperature electronics is shown in Extended Data Fig. \ref{fig:SetupDiagram}. Our transmitter consists of a 3D transmon qubit with resonance frequency $\omega_q/2\pi = 3.8945 $ GHz, $T_1=40 ~\mu s$ and $T_2^*=17 ~\mu s$. It is dispersively coupled to a superconducting aluminum cavity with a dispersive coupling parameter $\chi/2\pi \approx 0.2$ MHz. The relevant mode of the cavity has a resonance frequency of $\omega_\mathrm{cav.}/2\pi = 7.3918 $ GHz and linewidth $\kappa/2\pi=3.2$ MHz. To compensate for finite anharmonicity of our transmon qubit, we apply all pulses with a DRAG correction \cite{motzoi2009simple}. 

Our receiver consists of a Josephson parametric amplifier (JPA), a traveling wave parametric amplifier (TWPA), a high-electron mobility transistor (HEMT), a number of room-temperature amplifiers, and a field programmable gate array (FPGA) for feedback control. As we change the center frequency of the JPA's gain profile by rapidly modulating the pump frequency, we require a relatively high bandwidth. Counterintuitively, heterodyne is at a comparative disadvantage in our system, as the signal is detuned from the amplifier. To make sure that heterodyne and adaptivedyne operate with the same effective gain, we exploit the inherent trade-off between gain and bandwidth, operating the JPA at a relatively low gain of 6 dB, so that its bandwidth is $45$ MHz full-width half-maximum. The following TWPA has a gain of $15.3$ dB and an intrinsic GHz bandwidth. The combination maintains a high quantum efficiency of $\eta = 0.4$, which is standard for a JPA.

\textbf{Signal generation and detection}: As our signal consists of at most one photon, it is important to avoid spurious tones at the cavity frequency that could overwhelm it\cite{hacohen2016noncommuting}. We avoid this issue by detuning the local oscillator (LO) frequency $\omega_\mathrm{LO}$ below the cavity frequency by 105 MHz, chosen to avoid harmonics at important frequencies such as higher cavity modes. We then use heterodyne demodulation to down-convert the output signal back to lower frequencies for digitization and feedback. Qubit readout and bath engineering sideband tones are generated with a mixer by modulating the local oscillator at $105$ and $105+\Omega_\mathrm{R}/2\pi = 125$ MHz respectively. We also apply a cancellation tone to the output port of our cavity to cancel the sideband tone before it reaches the JPA. As the sideband tone is time-dependent to enable flat photon generation, it also contains power around its center frequency. To cancel it over a finite bandwidth, we use a room temperature cavity to match the dispersion of the cancellation tone line to the dispersion of the qubit cavity.

To generate the $2\omega_\mathrm{cav.}$ parametric pump for the JPA, we first double the local oscillator using a passive nonlinear doubler and then modulate this up-converted tone at $105\times 2=210$ MHz to generate a pump at $2\omega_\mathrm{cav.}$. This scheme again ensures that there are no unwanted coherent tones at $\omega_\mathrm{cav.}$. Modulation for cavity and qubit tones are generated by a Tektronix AWG5014C arbitrary waveform generator, while the JPA pump is modulated by the FPGA to enable fast feedback. All mixers used for signal generation are balanced to output only a single sideband.

\subsection{Single photon generation, mode shape control and feedback} \label{sec:DeriveAPM}

To generate our photon, we use the bath engineering scheme first described in \cite{murch2012cavity}. We work in a dressed frame of the qubit set by an applied Rabi drive of $\Omega_R/2\pi = 20$ MHz. As the qubit undergoes hundreds of coherent oscillations during photon emission, it is not possible to produce a phase-stable photon unless the Rabi frequency is actively stabilized\cite{hacohen2016noncommuting}. We maintain $\Omega_\mathrm{R}/2\pi$ to within $\pm 2$ kHz standard deviation by periodically running a pair of sequences in which we Rabi drive the qubit for $T_\mathrm{Rabi} \approx 4~\mu s$, measure the qubit state and then correct the Rabi drive amplitude based on the measurement result. 
By timing the measurement so that $\langle \sigma_z\rangle=0$ we ensure maximum sensitivity to small drifts in $\Omega_\mathrm{R}$.  We measure at two different points in time separated by half a period, so that a shift in $\Omega_\mathrm{R}$ leads to a differential shift in $\langle \sigma_z \rangle$ between the two time points, as opposed to readout drift, which shifts them in the same direction. The measurement time is chosen to satisfy a trade-off between sensitivity and maximum tolerable frequency drift before slipping to another period of the Rabi oscillation. If the qubit is measured $N$ times, then the uncertainty in the measured Rabi frequency is $1/(2\pi\sqrt{N}T_\mathrm{Rabi})$ while the maximum tolerable frequency drift is $1/8 T_\mathrm{Rabi}$.


The full experimental sequence is shown in Extended Data Fig. \ref{fig:StarkCal}A. We first focus on the generation of a photon with the desired flat mode shape, which requires a time-dependent cooling rate $\gamma(t)$. To implement these dynamics, we measure the induced cooling rate as a function of sideband amplitude as shown in Extended Data Fig. \ref{fig:StarkCal}B. The sideband also induces a Stark shift on the qubit, which in turn changes the Rabi drive amplitude that achieves $\Omega_\mathrm{R}/2\pi=20$ MHz.
These calibrations are plotted in Extended Data Fig. \ref{fig:StarkCal}C and D respectively. To emit the photon, we first ramp up the Rabi drive, and then apply the time-dependent sideband drive. As the sideband amplitude changes, we adjust the Rabi drive frequency and amplitude according to Extended Data Fig. \ref{fig:StarkCal} C and D respectively. The result is the flat photon shown in Fig. 1A, which is highly phase-stable. 

We use the full master equation to derive the required $\gamma(t)$, which also lets us calculate the optimal feedback strategy for a given photon mode shape. These results are also derived in \cite{pozza2015deterministic}, but we include a variant here for completeness. As photon loss does not affect the decay dynamics or the best estimate of the phase, we assume $\eta=1$ for this analysis. We also neglect other forms of decoherence, which have a negligible effect on the decay dynamics. This allows us to perform our computations with a pure state, so we begin with the unnormalized stochastic Schr\"odinger equation for an atom observed via homodyne detection\cite{Wiseman2009book}, which provides a state update from the acquisition of an infinitesimal amount of information via homodyne detection of the atom's spontaneous emission
\begin{smequation} \label{eq:SSE}
\frac{d}{dt}|\tilde{\psi}\rangle = \left[-\frac{1}{2} \gamma(t) \sigma^\dagger \sigma + \sqrt{\gamma(t)}e^{-i \phi(t)}\sigma V(t) \right]|\tilde{\psi}\rangle.
\end{smequation}
%
$V(t)$ is the measurement record, $\sigma=|-\rangle\langle+|$ and $|\tilde{\psi}\rangle$ is the unnormalized pure state describing the state of the atom\footnote{It is a slight abuse of notation to write Eq. \ref{eq:SSE} as a differential equation, as $V(t)$ is actually an unbounded stochastic quantity. Formally, it must be treated as an Ito integral\cite{Oksendal2003}, but this leads to no additional complications until the end of this section, at which point to address this issue directly}. This linear form of the equation is obtained in the same way that the state update under projective measurement may be made linear by dropping normalization \textit{i.e.} $|\tilde{\psi}\rangle \rightarrow |\Psi\rangle\langle\Psi|\tilde{\psi}\rangle$. See \cite{JacobsSteck2006} for a pedagogical introduction to continuous measurement. If we write $|\tilde{\psi}\rangle$ as $|\tilde{\psi}\rangle=c_-|-\rangle+c_+|+\rangle$, then the equation of motion for $c_+$ is
\begin{smequation}
\frac{dc_+}{dt} = -\frac{1}{2}\gamma(t)c_+ ~~\rightarrow~~ c_+(t) = c_+(0)e^{-\frac{1}{2}\int_0^t\gamma(s)ds}
\end{smequation}
where we have assumed that $\gamma(t<0)=0$. Recall that our system decays from $|+\rangle$ to $|-\rangle$. Although in general Eq. \ref{eq:SSE} does not preserve the norm of $|\tilde{\psi}\rangle$, one nevertheless derives the correct equation of motion for the average population from the above in the absence of measurement. The result is $d|c_+|^2/dt = \gamma(t)|c_+|^2$, which coincides with the expectation based on a standard rate equation for decay of the excited state population. We identify the mode shape with the instantaneous emitted intensity, assuming the atom was initialized with $c_+=1$, $c_-=0$
\begin{smequation}
u(t) \equiv \gamma(t)|c_+|^2 = \gamma(t) e^{-\int_0^t\gamma(s)ds}.
\end{smequation}
Notice that $u(t)$ integrates to 1 for any $\gamma(t)$. If we demand a flat mode shape so that $u(t)$ is constant, then $\gamma(t) = 1/(\tau-t)$, where $\tau=10\mu s$ parameterizes the photon's duration. As $\gamma(t)$ diverges at $t=\tau$, we set a maximum cooling rate of $1.4$ MHz and cool at this maximum rate for several microseconds longer than $\tau$, such that more than 99\% of the excited state population has decayed by $T=13\mu s$. The $\gamma(t)$ used experimentally is shown in Extended Data Fig. \ref{fig:StarkCal}A. The flat portion when $\gamma/2\pi=1.4$ MHz coincides with the portion of the photon that decays exponentially, as can be seen in Fig. 1C. 

Now that we have developed the necessary tools for emitting a flat photon, we derive the optimal feedback protocol given our photon. The equations of motion for $c_-$ determine the best estimate of the phase
\begin{smequation} \label{eq:EOMcminus}
\frac{dc_-}{dt} = c_+ \sqrt{\gamma(t)}e^{-i\phi(t)}V(t) ~~ \rightarrow ~~ c_-(t) = c_-(0) + c_+(0)\int_0^t e^{-i\phi(s)} \sqrt{u(s)} V(s) ds.
\end{smequation}
Notice the similarity between Eq. \ref{eq:EOMcminus} and Eq. 1 of the main text. 
For feedback, we wish to compute the best estimate of the atomic dipole phase at time $t$ assuming that the controller initially has no information about the phase. This best estimate coincides with the best estimate for the phase of the emitted photon after that time. To compute it, we note that the dynamics are trivial if the system is initialized in $|-\rangle$, so that the dipole phase evolution of the zero-knowledge mixed state $\rho_0 = (|-\rangle\langle-| + |+\rangle\langle+|)/2$ is entirely determined by the dynamics of the second term. Again taking $c_+=1$, $c_-=0$, the dipole phase is given by the relative complex phase between $c_+$ and $c_-^*$. As the complex phase of $c_+$ remains constant, the dipole moment phase is simply
\begin{smequation} \label{eq:RIntegralSM}
\theta(t) = \arg(R), ~~ R(t) \equiv \int_0^t e^{i\phi(s)} \sqrt{u(s)} V(s) ds
\end{smequation}
in agreement with Eq. 1
of the main text. In principle, Eq. 1
and the phase measurement condition $\phi(t) = \theta(t)+\pi/2$ define the optimal protocol. For ease of implementation, this protocol may further simplified by solving for the absolute value and complex argument of $R$ individually as follows. If the controller maintains the phase measurement condition, then we have $\exp(i\phi(s)) = i R/|R|$. Making this substitution and differentiating with respect to $t$ yields
\begin{smequation} \label{eq:dR}
dR = i \frac{R}{|R|}\sqrt{u(t)} V(t)dt.
\end{smequation}
To compute a differential equation for $|R|$, one must be aware that $V(t)$ is a random variable. $V(t)dt$ is unbounded, and the standard chain rule of differential calculus must be replaced with Ito's lemma, which looks like the chain rule but expanded to higher order like a Taylor series. As $V(t)$ is normalized to have a variance $dt$, $(V(t)dt)^2 = dt$ and we have
\begin{smequation}
d|R|^2 = u(t)dt ~~ \rightarrow ~~ |R(t)|^2 = \int_0^t u(s)ds.
\end{smequation}
Thus the time evolution of $|R|$ is deterministic. Substituting this solution into Eq. \ref{eq:dR} yields
\begin{smequation} 
dR = i R P(t) V(t)dt, ~~ P(t) \equiv \sqrt{\frac{u(t)}{\int_0^t u(s)ds}}.
\end{smequation}
Finally, we use Ito's lemma one more time to compute the differential increment of $\theta = \arg(R) = \mathrm{Im}[\log(R)]$
\begin{smequation} \label{eq:PofT}
d\theta = \mathrm{Im}(d\log(R)) = \mathrm{Im}\left[i P(t) V(t)dt + \frac{P(t)^2}{2}dt\right] = P(t) V(t) dt.
\end{smequation}
As $d\phi(t) = d\theta(t)$, Eq. \ref{eq:PofT} states that the instantaneous angular frequency of the measurement axis is proportional to the measurement outcome. Thus in the limit that the feedback delay is small, the process of computing a quantum trajectory and then calculating the optimal phase may be reduced to applying proportional feedback. We implement this feedback law in the FPGA, as described in section \ref{sec:FPGADetails}

\subsection{Experimental validation of quantum trajectories}

Each shot of the experiment consists of the following operations, shown in Extended Data Fig. \ref{fig:StarkCal}A. First, we perform a projective herald readout to check if the qubit is in the ground state, and then apply pulses to prepare any desired state $\rho(0)$ contained in the $\{|g\rangle,|e\rangle\}$ subspace. We then apply time-dependent cooling for fixed amount of time, stop cooling and then perform one of 7 tomography pulses and a strong readout. To enable checks between tomography and theory, we stop cooling at a time $t_f$ before the atom has fully decayed.

In post-processing, we simulate evolution of $\rho$ under the following stochastic master equation\cite{Wiseman2009book,campagne2016observing,JacobsSteck2006}
\begin{smequation} \label{eq:SME}
d\rho = \frac{\Gamma_{T2}}{2} \mathcal{D}[\sigma_z]\rho(t) dt + \gamma(t) \mathcal{D}[\sigma]\rho(t)dt + \sqrt{\gamma(t)\eta}\mathcal{H}[\sigma e^{-i\phi(t)}]\rho(t) dW(t)
\end{smequation}
\begin{smequation}
V(t)dt = \sqrt{\gamma(t)\eta}\langle \sigma e^{-i\phi(t)} + \sigma^\dagger e^{i\phi(t)}\rangle dt + dW(t)
\end{smequation}
where $\Gamma_{T2} = 60$ kHz is an empirically measured dephasing rate in the Rabi frame and $W(t)$ is a Wiener process. For $\eta=1$, $\Gamma_{T2}=0$, Eq. \ref{eq:SME} is equivalent to Eq. \ref{eq:SSE} except that the former preserves the norm of the state by keeping non-linear terms. In practice, we use a higher-order numerical method to propagate Eq. \ref{eq:SME} which guarantees positivity of the density matrix\cite{rouchon2015efficient}. To compare with experiment, we first compute $\rho(t_f)$ for each shot using the associated measurement record. We then find all shots in which the expectation value $\langle \sigma_x\rangle$, $\langle \sigma_y\rangle$ or $\langle \sigma_z\rangle$ are near a particular value and collect the corresponding tomography measurement outcomes. We compare $\langle \sigma_x\rangle$, $\langle \sigma_y\rangle$ and $\langle \sigma_z\rangle$ to the tomography data for adaptivedyne, heterodyne and homodyne detection for $t_f =$ 2, 4, 6, 8 and 10 $\mu$s. In each of these datasets, we prepare each of the 6 Clifford states, which amounts to a total of 90 validation datasets. In Extended Data Fig. \ref{fig:TrajValidation}, we plot a random sample of these datasets, chosen so that at least one of each dataset type is represented. The agreement between theory and experiment for the presented sample is representative of the entire dataset, which does not show any major deviations or apparent systematic errors. 



\subsection{Adaptive detector and feedback controller} \label{sec:FPGADetails}

As shown schematically in Extended Data Fig. \ref{fig:FPGACode}, we use an Innovative Integration X6-1000M FPGA board to control the flux pump tone for the JPA and to digitize the down-converted photon signal. The JPA pump is generated using two on-board 1 Gsample/s digital-to-analog converters (DACs) generating tones at 210 MHz, and an external I/Q mixer to perform single sideband modulation. The LO for this mixer is the frequency-doubled cavity LO, so that the JPA pump is twice the cavity frequency and phase locked to the photon. On the FPGA input side, the photon signal is down-converted to 105 MHz by an external mixer using the cavity LO, and is sampled by the FGPA board's analog-to-digital converters (ADCs) at 1 Gsample/s. Inside the FPGA, the signal is then demodulated to DC and filtered (Extended Data Fig. \ref{fig:FPGACode}A). During adaptive feedback, the pump tone is continuously detuned from the base frequency of 210 MHz by an amount proportional to the instantaneous value of the demodulated signal quadrature according to Eq. \ref{eq:PofT}. Since the JPA pump phase is continuously changing, we use the pump's instantaneous output phase delayed by the 374 ns electrical delay of the feedback cycle to determine the amplified quadrature of the input (Extended Data Fig. \ref{fig:FPGACode}D). To ensure that the correct quadrature is read out, we perform a separate calibration in which we sweep the pump frequency from 202 MHz to 218 MHz. We see less than a 0.5 degree variation between the pump frame delayed through the FGPA and the amplified quadrature measured from the JPA. We also ensure that the pump gain stays constant over this frequency band using the lookup table shown in Extended Data Fig. \ref{fig:GainCurve}, which determines the pump amplitude for a given pump frequency. Using the major axis variance of the squeezed vacuum ellipse as a proxy for gain, we intersperse the calibration of this lookup table throughout our measurements to compensate for slow changes in the pump chain, which occur due to room temperature amplifier drift. 

Alongside the feedback mode, the FGPA has two other modes: replay and heterodyne. In replay mode, the output pump waveform is identical to the waveform of the previous adaptive pump waveform; the JPA undergoes the same rotations as it had during adapting, but the pump phase is no longer correlated with the estimated photon phase. In heterodyne mode, the JPA pump frequency is detuned from 210 MHz by a constant heterodyne frequency, 0.5 MHz. During our measurements, the board cycles through adaptive, replay, and heterodyne, changing modes on each trigger of the AWG (Extended Data Fig. \ref{fig:FPGACode}B).




\subsection{Performance comparison between heterodyne and canonical phase measurement}

The data of Fig. 3
confirm that we implement an approximate canonical phase measurement. However, the relative comparison with heterodyne made in Fig. 4
leaves open two potential issues. Firstly, one may wonder if the chosen heterodyne frequency of $0.5$ MHz is sufficiently large to qualify as heterodyne detection, or if superior performance could be attained with a higher frequency. It can be shown that in the absence of feedback, the performance does not depend on full time-dependence of $\phi(t)$, but only on whether the angles are 'sampled uniformly' relative to the mode shape. More precisely, if the integral $\int_0^T e^{i\phi(t)} \sqrt{u(t)} dW$ leads to a rotationally uniform Gaussian probability distribution in the complex plane, then the system will perform at the heterodyne limit. As $u(t)$ is approximately constant, this condition is easy to satisfy. To confirm this analysis, we also numerically simulated the performance of heterodyne for various heterodyne frequencies. As can be seen in Extended Data Fig. \ref{fig:FvsfHet}, the intrinsic phase estimation efficiency\cite{pozza2015deterministic} saturates at the heterodyne limit once $\phi(t)$ makes a full revolution over the duration of the photon, indicating that the heterodyne frequency used experimentally more than suffices.

The second issue is whether the intrinsic quantum efficiency of the JPA depends on $\phi(t)$. If $d\phi/dt$ is large, then the gain of the amplifier decreases, so in general we expect the efficiency to decrease as we change the measurement basis more rapidly. The maximum frequency applied when adapting was $\sim 1$ MHz, which is much smaller than the 45 MHz bandwidth of the JPA. Furthermore, we digitally filter the signal with a $128$ ns exponential kernal before using it to apply feedback, so that the higher-order derivatives of $\phi$ are also limited. To confirm that these precautions maintain equal quantum efficiencies between adaptivedyne and heterodyne, we measure the signal-to-noise ratio (SNR) when a weak coherent tone is input to the amplifier. The SNR for heterodyne is $0.2\% \pm 0.3$\% larger than for adaptivedyne, which is negligible compared to the relative improvement observed.

\begin{smfigure}
{\includegraphics[width = 150mm]{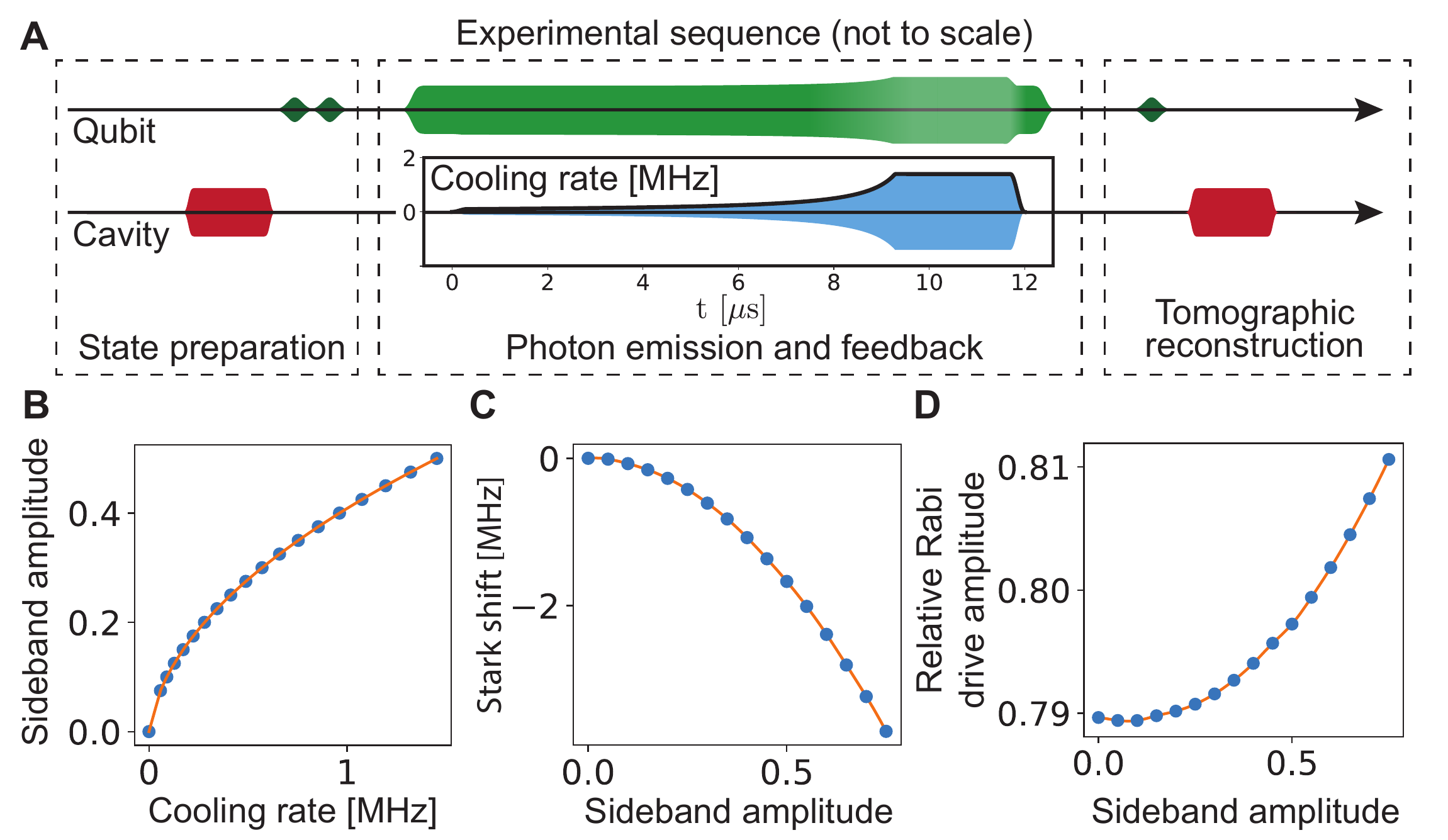}}
\caption{Pulse sequence and associated calibrations. (\textbf{A}) Pulse sequence of each shot of the experiment. Vertical axis represents the amplitude of each pulse, with the exception of the cooling sideband, in which the vertical axis is the quantitative cooling rate. Cavity-resonant pulses are shown in red, cavity sideband pulses in blue and qubit-resonant pulses in green. The discoloration and distortion of the central qubit pulse represent cooling-sideband-dependent frequency and amplitude modulation, as calibrated in (\textbf{C}) and (\textbf{D}) respectively. Effects are exaggerated for visual clarity. (\textbf{B}) Measurement of the bath engineering cooling rate versus sideband amplitude. (\textbf{C}) Measurement of the Stark shift induced by the sideband as a function of sideband amplitude (\textbf{D}) Sideband amplitude required to drive 20 MHz Rabi oscillations as a function of the sideband amplitude.}
\label{fig:StarkCal}
\end{smfigure}

\begin{smfigure}
{\includegraphics[width = 150mm]{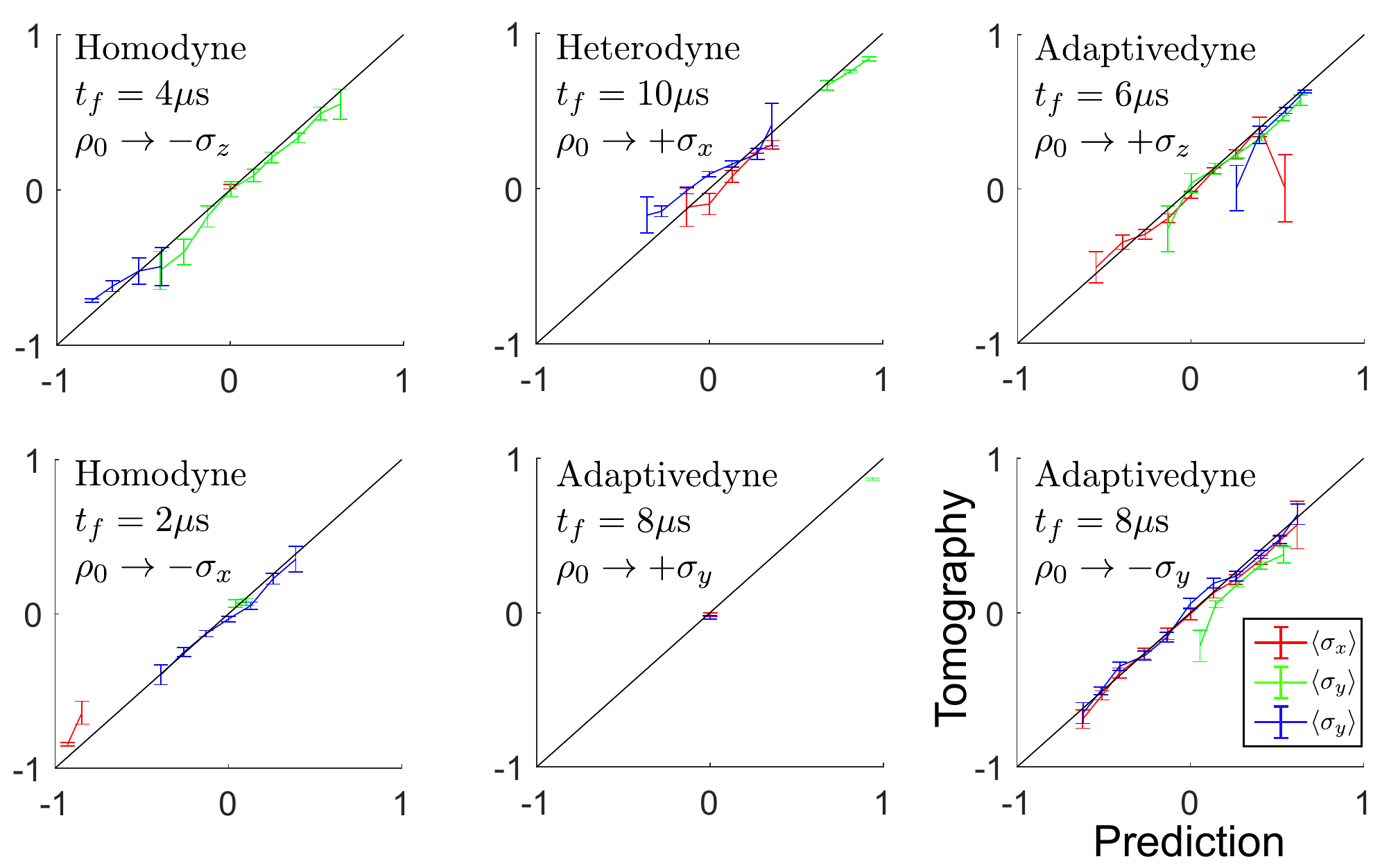}}
\caption{Trajectory validation for 6 randomly selected datasets. $\pm \sigma_i$ notation indicates that the eigenvector of $\sigma_i$ with associated eigenvalue $\pm 1$ was prepared as the initial state for the given dataset.}
\label{fig:TrajValidation}
\end{smfigure}

\begin{smfigure}
{\includegraphics[width = 85mm]{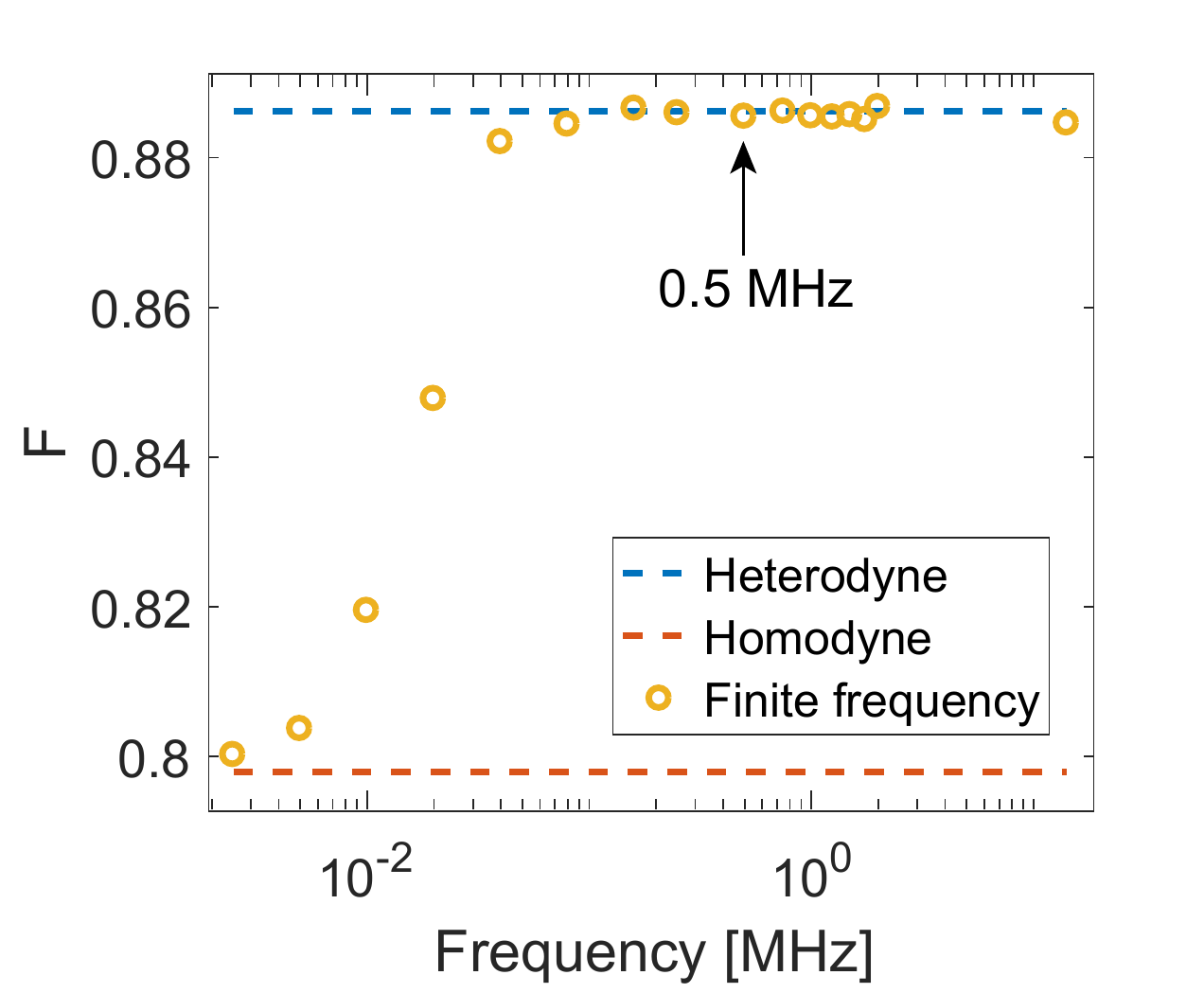}}
\caption{Intrinsic phase estimation efficiency $F$\cite{pozza2015deterministic}
as a function of heterodyne frequency. Efficiency saturates at the ideal theoretical value of $\sqrt{\pi}/2 \approx 0.886$ well below the heterodyne frequency used experimentally ($0.5$ MHz).
}
\label{fig:FvsfHet}
\end{smfigure}



\begin{smfigure}
{\includegraphics[width = 150mm]{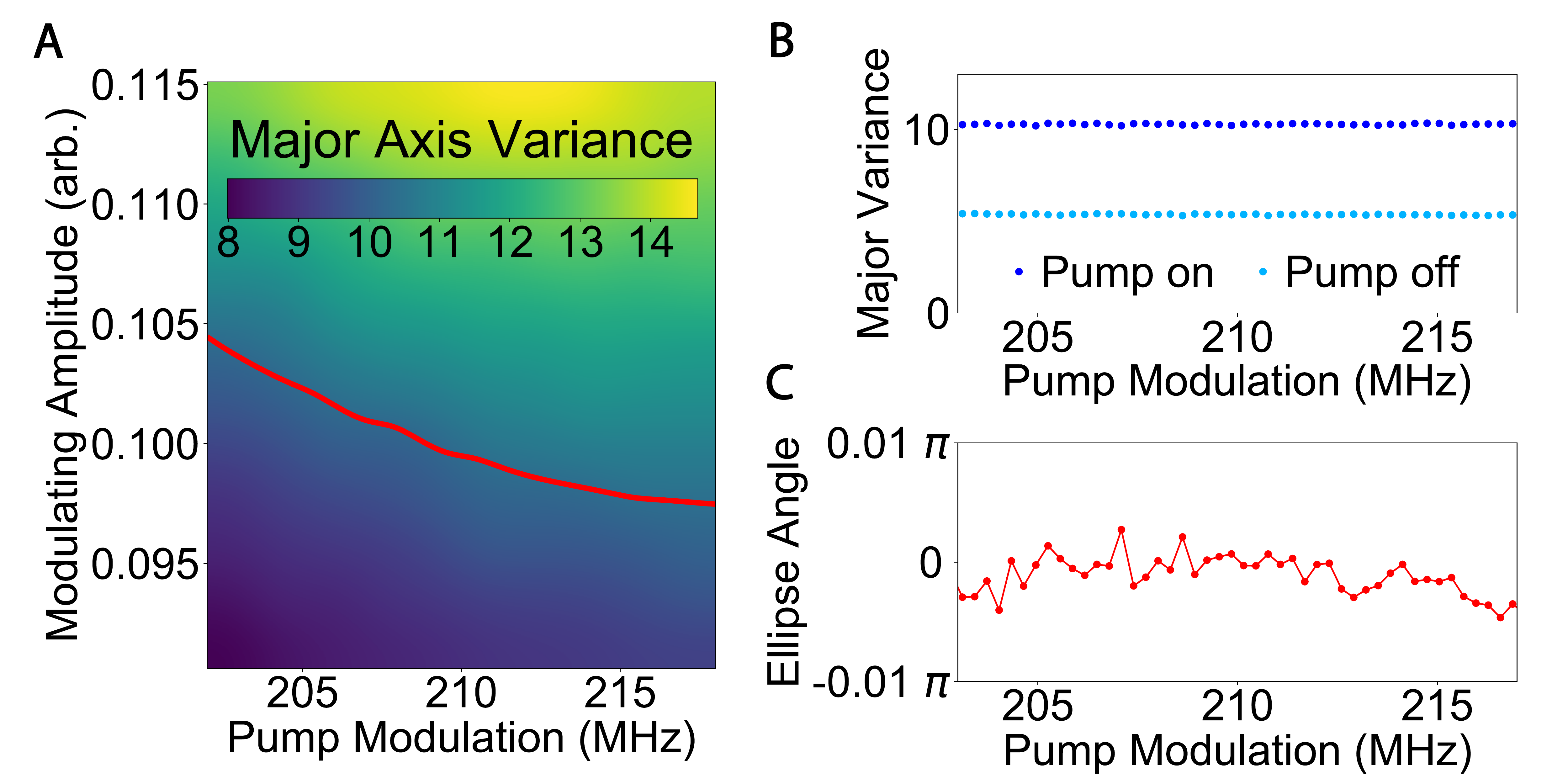}}
\caption{Gain calibration curve for the JPA.
(\textbf{A}) Major axis variance (arb. units) of amplified vacuum as a function of pump frequency and amplitude. The red line represents a contour of constant variance, the gain curve.
(\textbf{B}) Major axis variance of amplified vacuum along the gain curve with the JPA pump on as opposed to off.
(\textbf{C}) Angle in radians of the amplified vacuum's major axis along the gain curve. An electrical delay is calibrated to keep this curve flat across frequency.
}
\label{fig:GainCurve}
\end{smfigure}

\begin{smfigure}
{\includegraphics[width = 430pt]{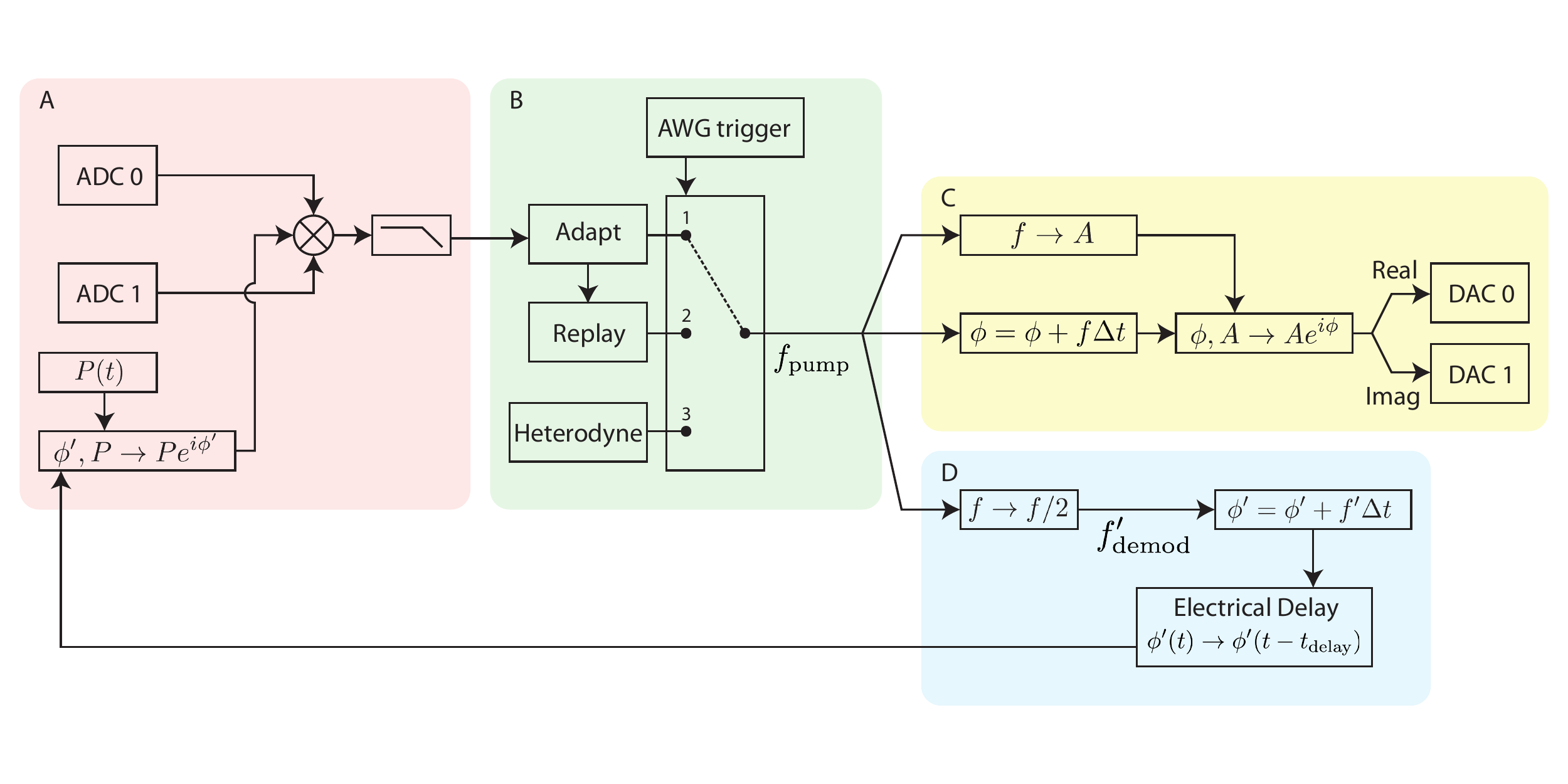}}
\caption{Internal logic block diagram for the FPGA.
(\textbf{A}) Photon signal demodulation. The demodulation phase $\phi'$ is determined by the JPA's pump phase. The demodulation amplitude $P(t)$ is given by Eq. \ref{eq:dR}.
(\textbf{B}) JPA pump frequency selection, advanced each trigger of the AWG: 1) adapting JPA frequency proportional to the incoming signal; 2) replaying JPA frequency from the previous AWG trigger; 3) heterodyning using a fixed JPA frequency.
(\textbf{C}) Instantaneous JPA frequency $f_\mathrm{pump}$ increments the JPA pump phase $\phi$ and determines the instantaneous amplitude $A$ through the gain calibration shown in Fig. \ref{fig:GainCurve}. The real and imaginary parts of this pump tone are sent to DAC0 and DAC1 respectively for single sideband modulation.
(\textbf{D}) The JPA pump frequency is halved, accumulated, and delayed to determine the demodulation phase $\phi'$.
}
\label{fig:FPGACode}
\end{smfigure}

\begin{smfigure}[hbtp]
\includegraphics[width=370pt]{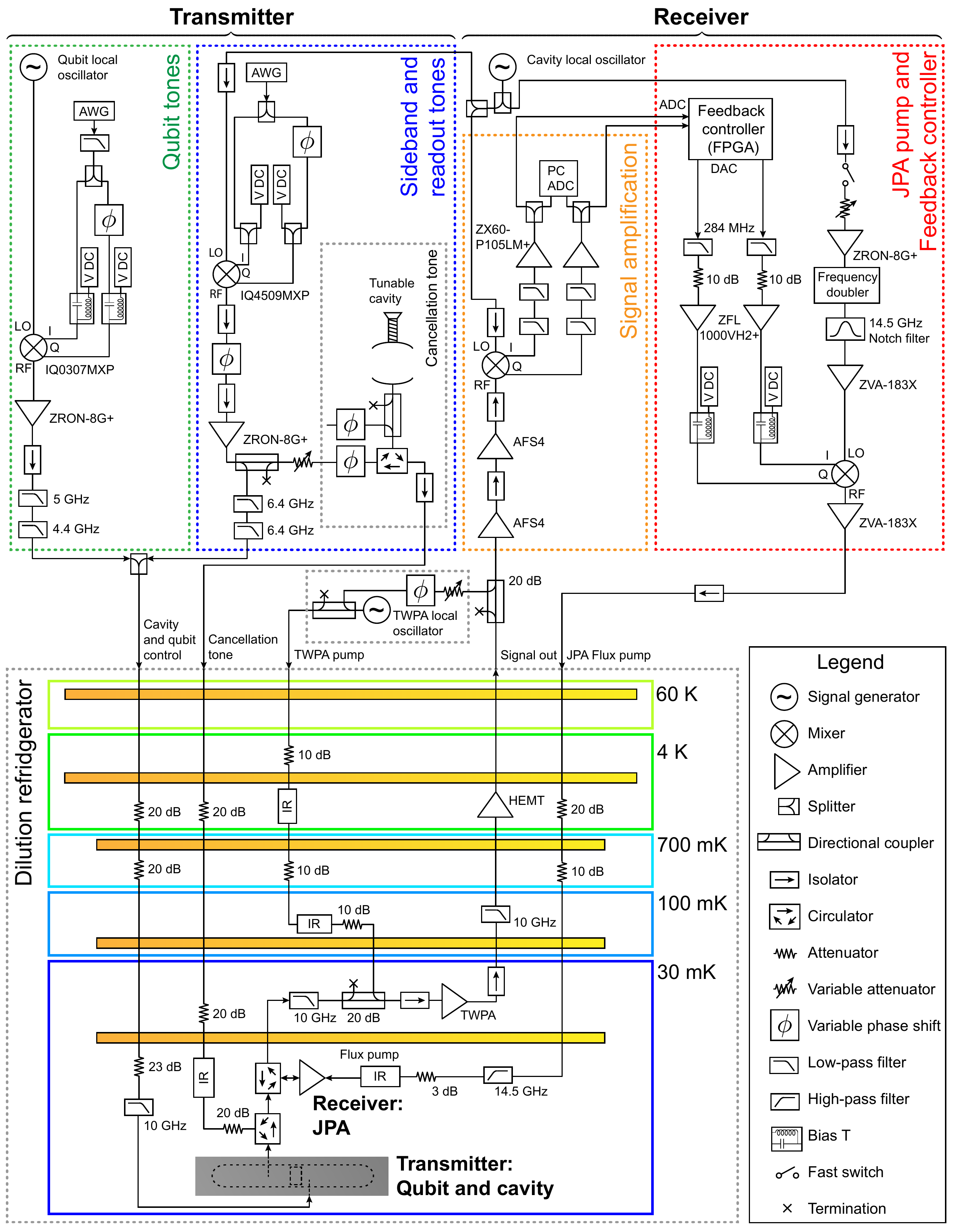}
\caption{Wiring diagram for the experiment, with relevant subsystems enclosed by gray dashed boxes. The JPA amplifies in reflection and is pumped and feedback-controlled via the flux pump line. IR filters are Eccorsorb or similar. The open wire in the cancellation tone is an intentional open circuit to tune the the cavity coupling.}
\label{fig:SetupDiagram}
\end{smfigure}

\end{methods}


\bibliography{APMbib1}



\begin{addendum}
 \item We thank Emmanuel Flurin, John Mark Kreikebaum and Vinay Ramasesh for assistance, and MIT Lincoln labs for fabrication of the traveling wave parametric amplifier. This work was supported by the Army Research Office under Grant \# W911NF15-1-0496. The effort of LM was supported by grants from the National Science Foundation Grant No. (1106400) and the Berkeley Fellowship for Graduate Study.
 \item[Competing Interests] The authors declare that they have no
competing financial interests.
 \item[Correspondence] Correspondence and requests for materials
should be addressed to L. S. Martin (email: leigh@berkeley.edu).
\end{addendum}


\end{document}